\documentclass[twocolumn,showpacs,aps,prd,superscriptaddress]{revtex4}

\usepackage{graphicx}
\usepackage{dcolumn}
\usepackage{amsmath}
\usepackage{epsfig}

\RequirePackage{xspace}

\usepackage{relsize}
\def\babar{\mbox{\slshape B\kern-0.1em{\smaller A}\kern-0.1em
    B\kern-0.1em{\smaller A\kern-0.2em R}}}

\def\epem       {\ensuremath{e^+e^-}\xspace}

\def\ellp       {\ensuremath{\ell^+}\xspace}

\def\nul        {\ensuremath{\nu_\ell}\xspace}

\def\piz   {\ensuremath{\pi^0}\xspace}

\def\pip   {\ensuremath{\pi^+}\xspace}
\def\pim   {\ensuremath{\pi^-}\xspace}

\def\Kbar  {\kern 0.2em\overline{\kern -0.2em K}{}\xspace}

\def\Kz    {\ensuremath{K^0}\xspace}
\def\Kzb   {\ensuremath{\Kbar^0}\xspace}
\def\KzKzb {\ensuremath{\Kz \kern -0.16em \Kzb}\xspace}
\def\Kp    {\ensuremath{K^+}\xspace}
\def\Km    {\ensuremath{K^-}\xspace}

\def\KpKm  {\ensuremath{\Kp \kern -0.16em \Km}\xspace}

\def\Dbar    {\kern 0.2em\overline{\kern -0.2em D}{}\xspace}

\def\Dz      {\ensuremath{D^0}\xspace}
\def\Dzb     {\ensuremath{\Dbar^0}\xspace}
\def\DzDzb   {\ensuremath{\Dz {\kern -0.16em \Dzb}}\xspace}
\def\Dp      {\ensuremath{D^+}\xspace}
\def\Dm      {\ensuremath{D^-}\xspace}

\def\DpDm    {\ensuremath{\Dp {\kern -0.16em \Dm}}\xspace}
\def\Dstar   {\ensuremath{D^*}\xspace}

\def\Dstarp  {\ensuremath{D^{*+}}\xspace}
\def\Dstarm  {\ensuremath{D^{*-}}\xspace}

\newcommand{\dsp}{\ensuremath{\Dstarp}\xspace}
\newcommand{\dsm}{\ensuremath{\Dstarm}\xspace}

\def\B       {\ensuremath{B}\xspace}
\def\Bbar    {\kern 0.18em\overline{\kern -0.18em B}{}\xspace}

\def\BB      {\ensuremath{B\Bbar}\xspace} 
\def\Bz      {\ensuremath{B^0}\xspace}
\def\Bzb     {\ensuremath{\Bbar^0}\xspace}
\def\BzBzb   {\ensuremath{\Bz {\kern -0.16em \Bzb}}\xspace}
\def\Bu      {\ensuremath{B^+}\xspace}
\def\Bub     {\ensuremath{B^-}\xspace}

\def\BpBm    {\ensuremath{\Bu {\kern -0.16em \Bub}}\xspace}

\def\BorBbar    {\kern 0.18em\optbar{\kern -0.18em B}{}\xspace}
\def\DorDbar    {\kern 0.18em\optbar{\kern -0.18em D}{}\xspace}
\def\KorKbar    {\kern 0.18em\optbar{\kern -0.18em K}{}\xspace}

\mathchardef\Upsilon="7107
\def\Y#1S{\ensuremath{\Upsilon{(#1S)}}\xspace}

\def\FourS {\Y4S}

\mathchardef\Deltares="7101
\mathchardef\Xi="7104
\mathchardef\Lambda="7103
\mathchardef\Sigma="7106
\mathchardef\Omega="710A

\def\Deltabar{\kern 0.25em\overline{\kern -0.25em \Deltares}{}\xspace}
\def\Lbar{\kern 0.2em\overline{\kern -0.2em\Lambda\kern 0.05em}\kern-0.05em{}\xspace}
\def\Sigbar{\kern 0.2em\overline{\kern -0.2em \Sigma}{}\xspace}
\def\Xibar{\kern 0.2em\overline{\kern -0.2em \Xi}{}\xspace}
\def\Obar{\kern 0.2em\overline{\kern -0.2em \Omega}{}\xspace}
\def\Nbar{\kern 0.2em\overline{\kern -0.2em N}{}\xspace}
\def\Xb{\kern 0.2em\overline{\kern -0.2em X}{}\xspace}

\newcommand{\tev}{\ensuremath{\mathrm{\,Te\kern -0.1em V}}\xspace}
\newcommand{\gev}{\ensuremath{\mathrm{\,Ge\kern -0.1em V}}\xspace}
\newcommand{\mev}{\ensuremath{\mathrm{\,Me\kern -0.1em V}}\xspace}
\newcommand{\kev}{\ensuremath{\mathrm{\,ke\kern -0.1em V}}\xspace}
\newcommand{\ev}{\ensuremath{\mathrm{\,e\kern -0.1em V}}\xspace}
\newcommand{\gevc}{\ensuremath{{\mathrm{\,Ge\kern -0.1em V\!/}c}}\xspace}
\newcommand{\mevc}{\ensuremath{{\mathrm{\,Me\kern -0.1em V\!/}c}}\xspace}
\newcommand{\gevcc}{\ensuremath{{\mathrm{\,Ge\kern -0.1em V\!/}c^2}}\xspace}
\newcommand{\mevcc}{\ensuremath{{\mathrm{\,Me\kern -0.1em V\!/}c^2}}\xspace}

\def\invfb   {\ensuremath{\mbox{\,fb}^{-1}}\xspace}

\def\mus  {\ensuremath{\rm \,\mus}\xspace}

\def\mus        {\ensuremath{\,\mu{\rm s}}\xspace}    

\def\ra                 {\ensuremath{\rightarrow}\xspace}
\def\to                 {\ensuremath{\rightarrow}\xspace}

\def\pep2{PEP-II}

\def\gsim{{~\raise.15em\hbox{$>$}\kern-.85em
          \lower.35em\hbox{$\sim$}~}\xspace}
\def\lsim{{~\raise.15em\hbox{$<$}\kern-.85em
          \lower.35em\hbox{$\sim$}~}\xspace}

\def\Vcb  {\ensuremath{|V_{cb}|}\xspace}

\xspace

\newcommand{\jrmp}      [1]  {{Rev.\ Mod.\ Phys.\ {\bf #1}}}  

\def\jetset74   {\mbox{\tt Jetset \hspace{-0.5em}7.\hspace{-0.2em}4}\xspace}

\newcommand{\bi}{\begin{itemize}}
\newcommand{\ei}{\end{itemize}}
\newcommand{\ben}{\begin{enumerate}}
\newcommand{\een}{\end{enumerate}}
\newcommand{\bc}{\begin{center}}
\newcommand{\ec}{\end{center}}
\newcommand{\bt}{\begin{table}}
\newcommand{\et}{\end{table}}
\newcommand{\be}{\begin{equation}}
\newcommand{\eeq}{\end{equation}}
\newcommand{\ba}{\begin{eqnarray}}
\newcommand{\ea}{\end{eqnarray}}

\newcommand{\la}{\ifmmode {\leftarrow} \else {$\leftarrow$}\fi}
\newcommand{\Ra}{\ifmmode {\Rightarrow} \else {$\Rightarrow$}\fi}
\newcommand{\La}{\ifmmode {\Leftarrow} \else {$\Leftarrow$}\fi}
\newcommand{\Lra}{\ifmmode {\Longrightarrow} \else {$\Longrightarrow$}\fi}
\newcommand{\Lla}{\ifmmode {\Longleftarrow} \else {$\Longleftarrow$\fi}}
\newcommand{\Llra}{\ifmmode {\Longleftrightarrow} \else {$\Longleftrightarrow$\fi}}
\newcommand{\Lk}{\ifmmode {{\cal L}} \else {${\cal L}$}\fi}
\newcommand{\Wt}{\ifmmode {{\cal W}} \else {${\cal W}$}\fi}
\newcommand{\Br}{\ifmmode {{\cal B}} \else {${\cal B}$}\fi}
\newcommand{\N}{\ifmmode {{\cal N}} \else {${\cal N}$}\fi}
\newcommand{\G}{\ifmmode {{\cal G}} \else {${\cal G}$}\fi}
\newcommand{\E}{\ifmmode {{\cal E}} \else {${\cal E}$}\fi}

\newcommand{\tBz}{\ifmmode {\tau_{\Bz}} \else {$\tau_{\Bz}$ }\fi }
\newcommand{\tBp}{\ifmmode {\tau_{\Bu}} \else {$\tau_{\Bu}$ }\fi }
\def\nubar    {\kern 0.18em\overline{\kern -0.18em \nu}{}\xspace}

\newcommand{\psoft}{\ifmmode {\pi_s^-} \else {$\pi_s^-$}\fi }
\newcommand{\dm}{\ifmmode {\Delta m} \else {$\Delta m$}\fi}
\newcommand{\plab}{\ifmmode{p} \else {$p$}\fi}
\newcommand{\ks}{\ifmmode{k^*} \else {$k^*$}\fi}
\newcommand{\om}{\ifmmode{w} \else {$w$}\fi}
\newcommand{\omt} {\ifmmode {\tilde{w}} \else {$\tilde{w}$} \fi}
\newcommand{\mnusq}{\ifmmode{{M_\nu}^2} \else {${M_{\nu}}^2$}\fi} 
\newcommand{\DTau}{\ifmmode {\Delta \tau} \else {$\Delta \tau$}\fi}
\newcommand{\ggcc}{\ifmmode {GeV^2/c^4} \else {$GeV^2/c^4$}\fi}
\newcommand{\TBY}{\ifmmode{\theta_{B, D^*\ell}} \else {$\theta_{B, D^*\ell}$} \fi}
\newcommand{\Aone}{\ifmmode {{\cal A}_1} \else {${\cal A}_1$}\fi}
\newcommand{\Atwo}{\ifmmode {{\cal A}_2} \else {${\cal A}_2$}\fi}
\newcommand{\Rone}{\ifmmode {{\cal R}_1} \else {${\cal R}_1$}\fi}
\newcommand{\Rtwo}{\ifmmode {{\cal R}_2} \else {${\cal R}_2$}\fi}
\newcommand{\rha}{\ifmmode{\mbox{\rho^2_{{\cal A}_1}}} \else {\mbox{$\rho^2_{{\cal A}_1}$}}\fi}
\def\BpBm {\ensuremath{B^+ {\kern -0.16em \Bub}}}

\def\ctl        {\ensuremath{{\cos\theta_\ell}}\xspace}
\def\ctv        {\ensuremath{{\cos\theta_V}}\xspace}
\def\angchi     {\ensuremath{{\chi}}\xspace}

\def\thetal        {\ensuremath{{\theta_\ell}}\xspace}
\def\thetav        {\ensuremath{{\theta_V}}\xspace}

\def\ctl        {\ensuremath{{\cos\theta_\ell}}\xspace}
\def\ctv        {\ensuremath{{\cos\theta_V}}\xspace}

\newcommand{\Dsl}{\ensuremath{\Dstar\ell}\xspace}

\newcommand{\BztoDslnu}{\ensuremath{\Bz\rightarrow\Dstarm\ell^{+}\nul}\xspace}

\newcommand{\BABARPubYear}    {07}
\newcommand{\BABARPubNumber}  {008}

\newcommand{\SLACPubNumber} {12511}

\def\figurebox#1#2#3{%
    \def\arg{#3}%
    \ifx\arg\empty
    {\hfill\vbox{\hsize#2\hrule\hbox to #2{\vrule\hfill\vbox to #1{\hsize#2\vfill}\vrule}\hrule}\hfill}%
    \else
    {\hfill\epsfbox{#3}\hfill}%
    \fi}

\begin{document}

\preprint{\babar-PUB-\BABARPubYear/\BABARPubNumber} 
\preprint{SLAC-PUB-\SLACPubNumber} 

\begin{flushleft}
\babar-PUB-\BABARPubYear/\BABARPubNumber\\
SLAC-PUB-\SLACPubNumber\\
\end{flushleft}

\vspace*{-0.1cm}
\title{
{\large \bf Determination of the Form Factors for the Decay {\boldmath \BztoDslnu} 
and of the CKM Matrix Element {\boldmath \Vcb}}}

%
\author{B.~Aubert}
\author{M.~Bona}
\author{D.~Boutigny}
\author{Y.~Karyotakis}
\author{J.~P.~Lees}
\author{V.~Poireau}
\author{X.~Prudent}
\author{V.~Tisserand}
\author{A.~Zghiche}
\affiliation{Laboratoire de Physique des Particules, IN2P3/CNRS et Universit\'e de Savoie, F-74941 Annecy-Le-Vieux, France }
\author{J.~Garra~Tico}
\author{E.~Grauges}
\affiliation{Universitat de Barcelona, Facultat de Fisica, Departament ECM, E-08028 Barcelona, Spain }
% ...
\collaboration{The \babar\ Collaboration}
\noaffiliation

\date{\today}

\begin{abstract}
We present a combined measurement of the Cabibbo-Kobayashi-Maskawa
matrix element $|V_{cb}|$ and of the parameters $\rho^2$, $R_1(1)$,
and $R_2(1)$, which fully characterize the form factors for the $B^0
\rightarrow D^{*-}\ell^{+}\nu_\ell$ decay in the framework of
heavy-quark effective theory.
The results, based on a selected sample of about 52,800 $B^0
\rightarrow D^{*-}\ell^{+}\nu_\ell$ decays, recorded by the
\mbox{\slshape B\kern-0.1em{\smaller A}\kern-0.1em
  B\kern-0.1em{\smaller A\kern-0.2em R}} detector, are $\rho^2 = 1.157
\pm 0.094 \pm 0.027$, $R_1(1) = 1.327 \pm 0.131 \pm 0.043$, $R_2(1) =
0.859 \pm 0.077 \pm 0.021$, and $\mathcal{F}(1)|V_{cb}| = (34.7 \pm
0.4 \pm 1.0) \times 10^{-3}$.  The first error is the statistical and
the second is the systematic uncertainty.  Combining these
measurements with the previous \mbox{\slshape B\kern-0.1em{\smaller
    A}\kern-0.1em B\kern-0.1em{\smaller A\kern-0.2em R}} measurement
of the form factors, which employs a different fit technique on a
partial sample of the data, we improve the statistical precision of
the result, $\rho^2 = 1.191 \pm 0.048 \pm 0.028, R_1(1) = 1.429 \pm
0.061 \pm 0.044, R_2(1) = 0.827 \pm 0.038 \pm 0.022, $ and $
\mathcal{F}(1)|V_{cb}| = (34.4 \pm 0.3 \pm 1.1) \times 10^{-3}.$ Using
lattice calculations for the axial form factor $\mathcal{F}(1)$, we
extract $|V_{cb}| =(37.4 \pm 0.3 \pm 1.2 \pm ^{1.2}_{1.4} ) \times
10^{-3}$, where the third error is due to the uncertainty in
$\mathcal{F}(1)$.  We also present a measurement of the exclusive
branching fraction, ${\cal B} = (4.69 \pm 0.04 \pm 0.34)\%$.
\end{abstract}


\maketitle
\section{INTRODUCTION}
\label{sec:intro}

The study of the semileptonic decay \BztoDslnu~\cite{footnote1} is
interesting in many respects. 
In the standard model,
the rate of this weak decay is proportional to the
square of the Cabibbo-Kobayashi-Maskawa (CKM) matrix element
$V_{cb}$, which is a measure of the weak coupling of the $b$ to the $c$
quark. 
This decay is also influenced by strong interactions. Their effect can
be parameterized by two axial form factors $A_1$ and $A_2$, and
one vector form factor $V$, each of which depends on the momentum
transfer squared $q^2$ of the $B$ meson to the \Dstar\ meson. 
The form of this dependence is not known {\it a priori}. 
In the framework of heavy-quark effective field theory (HQET)~\cite{ref:NeubertPhysReport,ref:Rich}, these three form
factors are related to each other through heavy quark symmetry (HQS),
but HQET leaves three free parameters, which must be determined by
experiment. 

The extraction of \Vcb\ relies on the measurement of differential
decay rates.  HQS predicts the normalization of the decay rate at the
maximum $q^2$, and \Vcb\ is determined from an extrapolation of the
form factors to this value.  The precise determination requires
corrections to the HQS prediction for the normalization, as well as a
measurement of the variation of the form factors near the maximum
$q^2$, where the decay rate goes to zero as the phase space vanishes.

Several experiments have measured \Vcb based on studies of the
differential decay width 
for \BztoDslnu
decays~\cite{ref:ARGUS,ref:BELLE,ref:ALEPH1,ref:ALEPH2,ref:OPAL,ref:DELPHI,ref:bad776}.
These analyses of the one-dimensional 
differential decay rate resulted in the measurement of only one of the 
form-factor parameters, 
and they relied on a measurement by the CLEO
Collaboration~\cite{ref:CLEOff} for the other two.
The uncertainty in these two parameters introduces the largest
systematic uncertainty in all previous measurements of \Vcb\ using this method.

Furthermore, for measurements of $|V_{ub}|$, based on both inclusive and exclusive
$B\ra X_u\ell\nu$ decays, improved knowledge of all form factors is
important to correctly describe the dominant $B\ra X_c \ell\nu$
background.

In this paper, we present a simultaneous measurement of $\Vcb$, of the
branching fraction for \BztoDslnu, and of the three form-factor
parameters, based on measurements of three one-dimensional decay
distributions.
Thus we extend the earlier \babar\ measurement~\cite{ref:bad776} where
$F(1)\Vcb$ and one of the form-factor parameters are measured,
fully accounting for correlations between these one-dimensional
distributions. 

We combine the results of this analysis with another \babar\
measurement of the form factors~\cite{ref:bad1224}, which employs a
fit to the full four-dimensional decay distribution on a partial sample
of the data.

The \Dstarm candidates are reconstructed from the \Dstarm\ra\Dzb\pim\
decays and the \Dzb\ mesons are reconstructed in three different
decay modes, $\Kp\pim$, $\Kp\pim\pip\pim$, and
$\Kp\pim\piz$. Electrons or muons are paired with the \Dstarm\
to form signal candidates.  The large data sample permits a 
precise determination of the background contributions, 
largely based on data, and thus results in smaller experimental uncertainties.

This leads to a further reduction of the form-factor errors.

\section{FORMALISM}
\label{sec:formalism}

\subsection{Kinematic variables}

\begin{figure}[ht]
\begin{center}
  \scalebox{0.9}{\includegraphics{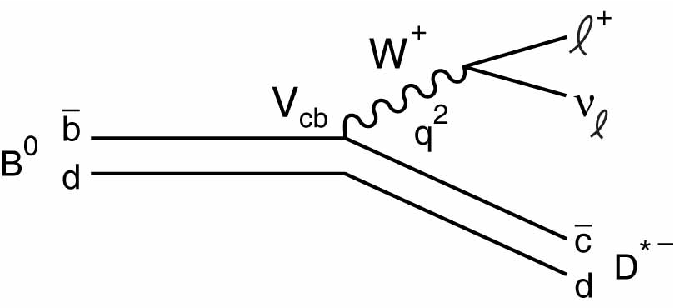}}
\end{center}
\vspace{-0.4cm}
\caption{Quark-level Feynman diagram for the decay \BztoDslnu.}
\label{fig:quarkleveldiag}
\end{figure}

The lowest-order quark-level diagram for the decay \BztoDslnu is shown in Figure~\ref{fig:quarkleveldiag}.
This decay is completely characterized by four variables, namely
three angles and the Lorentz-invariant variable $w$, which is linearly related to $q^2$ and defined  as

\begin{equation}
\label{eq:wdef}
w 
\equiv \frac{P_B \cdot P_{D^*}}{ m_B m_{D^*}} =  \frac{m_B^2 + m_{D^*}^2 -q^2} {2 m_B m_{D^*}},
\end{equation} 
where $m_B$ and $m_{D^*}$ are the masses of the $B$ and the $D^*$
mesons (2.010 and 5.2794 GeV respectively~\cite{ref:pdg07}), and $P_B$
and $P_{D^*}$ are their four-momenta.
In the $B$ rest frame the
expression for $w$ reduces to the Lorentz boost $\gamma_{D^*} =
E_{D^*}/m_{D^*} $.

The ranges of $w$ and $q^2$ are restricted by the kinematics of the
decay, with $q^2 = 0$ corresponding to
\begin{equation}
w_{max}=\frac{m_B^2 + m_{D^*}^2}{2m_B m_{D^*}} = 1.504
\end{equation} 
and $w_{min}=1$ corresponding to
\begin{equation}
q_{max}^2=(m_B-m_{D^*})^2 = 10.69~(\gev)^2.
\end{equation}
 
\begin{figure}[ht]
\begin{center}
  \scalebox{0.7}{\includegraphics{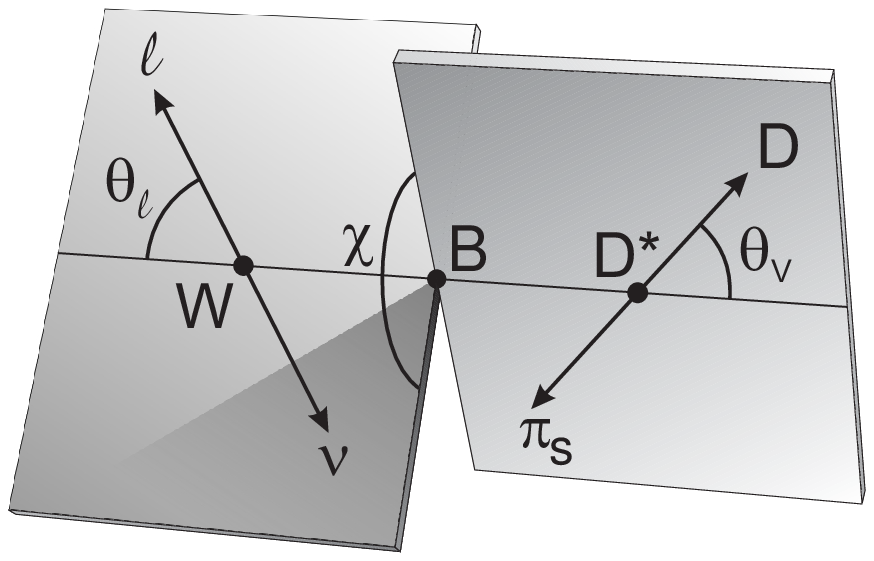}}
\end{center}
\vspace{-0.4cm}
\caption{Definition of the angles $\thetal$, $\thetav$, and $\angchi$
  for the \BztoDslnu decay, mediated by a vector boson $W$; $\pi_s$
  refers to the low momentum pion from the decay $D^{*-} \rightarrow
  \Dzb \pi^-_s$.}
\label{fig:mesonleveldiag}
\end{figure}

The three angular variables, shown in Figure~\ref{fig:mesonleveldiag},
are as follows:

\begin{itemize} 
  
\item{ $\thetal$, the angle between the direction of the lepton
    in the virtual $W$ rest frame and the direction of the $W$ in the
    $B$ rest frame; }
  
\item{ $\thetav$, the angle between the direction of the $D$ in the
    $D^*$ rest frame and the direction of the $D^*$ in the $B$ rest
    frame; }
  
\item{$\angchi$, the angle between the plane formed by the
    $D^{*}$ and the plane formed by the $W$ decay. }

\end{itemize} 

\subsection{ Four-dimensional decay distribution}

The Lorentz structure of the \BztoDslnu decay amplitude can be
expressed in terms of three helicity amplitudes ($H_{+}$, $H_{-}$, and
$H_{0}$), which correspond to the three polarization states of the
$D^*$, two transverse and one longitudinal.  For low-mass leptons,
these amplitudes are expressed in terms of the three functions
$h_{A_1}(w)$, $R_1(w)$, and
$R_2(w)$~\cite{ref:NeubertPhysReport,ref:Rich}:
\begin{equation}
H_i(w)=m_B\dfrac{R^*(1-r^2)(w+1)}{2\sqrt{1-2wr+r^2}}h_{A_1}(w)\tilde{H}_i(w),
\label{eq:HandHtilde}
\end{equation}
where
\begin{eqnarray}
\tilde{H}_{\mp} & = & \dfrac{\sqrt{1-2wr+r^2}\biggl(1\pm\sqrt{\dfrac{w-1}{w+1}}
R_1(w)\biggr)}{1-r}, \\
\tilde{H}_0 & = & 1+\dfrac{(w-1)(1-R_2(w))}{1-r},
\label{eq:HtildeandFF}
\end{eqnarray}
with
$R^*=(2\sqrt{m_B m_{\Dstar}})/(m_B+m_{\Dstar})$ and
$r=m_{\Dstar}/m_B$. \\
The functions $R_1(w)$ and $R_2(w)$ are defined in terms of the axial
and vector form factors as,
\begin{equation}
A_2(w) \equiv \frac {R_2(w)}{R^{*2}} \frac {2}{w+1} A_1(w),
\end{equation}
\begin{equation}
V(w) \equiv \frac {R_1(w)}{R^{*2}} \frac {2}{w+1} A_1(w).
\end{equation}
By convention, the function $h_{A_1}(w)$ is defined as,
\begin{equation}
h_{A_1}(w) \equiv \frac {1}{R^*} \frac {2}{w+1} A_1(w).
\end{equation}
For $w \to 1$, the axial form factor $A_1(w)$ dominates, and in the limit of infinite $b-$ and $c-$quark masses, a single form factor describes
the decay, the so-called Isgur-Wise function~\cite{ref:isgurwise}.

The fully differential decay rate in terms of the three helicity
amplitudes is
\begin{equation}
\label{eq:totaldiffdecaywidth}
\begin{split}
& \dfrac{\mathrm{d^4\Gamma}(\BztoDslnu)}{\mathrm{d}w
  \mathrm{d\cos\theta_{\ell}} \mathrm{d\cos\theta_{V}} \mathrm{d\chi}}
= \dfrac{6m_B~m_{\Dstar}^2}{8(4\pi)^{4}} \\
& \times \sqrt{w^2-1}(1-2wr+r^2) G_{F}^{2}|V_{cb}|^{2}\\
& \times \big\{ (1-\cos\theta_{\ell})^{2}\sin^{2}\theta_{V} H^2_{+}(w)
  \biggr. \\  
& + (1+\cos\theta_{\ell})^{2}\sin^{2}\theta_{V} H^2_{-}(w)  \\ 
& + 4\sin^{2}\theta_{\ell}\cos^{2}\theta_{V} H^2_{0}(w) \\ 
& - 2\sin^2\theta_{\ell}\sin^2\theta_{V}\cos 2\chi H_{+}(w)H_{-}(w) \\     
& - 4\sin\theta_{\ell}(1-\cos\theta_{\ell})\sin\theta_{V}\cos\theta_{V}\cos\chi \\ 
& \times H_{+}(w)H_{0}(w) \\  
& + \biggl. 4\sin\theta_{\ell}(1+\cos\theta_{\ell})\sin\theta_{V}\cos\theta_{V}\cos\chi \\  
& \times H_{-}(w)H_{0}(w)
\big\}.
\end{split}
\end{equation}
By integrating this decay rate over all but one of the four variables,
$w$, $\ctl$, $\ctv$, or $\chi$, we obtain the four one-dimensional
decay distributions from which we extract the form factors. 
The differential decay rate as a function of $w$ is
\begin{equation}
\label{eq:dgamma}
\frac{{\rm d}\Gamma}{{\rm d}\om}
= \frac{G^2_F}{48\pi^3} m_{\dsp}^3 \big[ m_{\Bzb}-m_{\dsp}\big]^2 
{\cal G}(w){\cal F}^2(w) \Vcb^2 ,
\end{equation}
where  
\begin{eqnarray*}
{\cal F}^2(w){\cal G}(w) = h_{A_1}^2(\om) \sqrt{\om-1} (\om+1)^2 \left\{ 2 \left[\dfrac{1-2\om r+r^2}{(1-r)^2}\right] \right. \\
\left. \times \left[1+R_1(\om)^2 \dfrac{\om-1}{\om+1}\right] + \left[1+(1-R_2(\om)) \dfrac{\om-1}{1-r}\right]^2 \right\},
\end{eqnarray*}
and ${\cal G}(\om)$ is a known phase space factor,
\begin{equation*}
{\cal G}(\om)= \sqrt{\om^2-1}(\om+1)^2 \left[1 + 4 \frac{\om}{\om+1} \frac{1-2\om r+r^2}{(1-r)^2}\right].
\end{equation*}

It is important to note that $h_{A_{1}}(1) \equiv \mathcal{F}(1)$
corresponds to the Isgur-Wise function~\cite{ref:isgurwise} at $w
=1$. In the infinite quark-mass limit, the HQS normalization gives
$\mathcal{F}(1)=1$. Corrections to this HQS prediction have been
calculated in the framework of lattice QCD.
A recent calculation, performed in a quenched approximation, predicts (including a QED correction of 0.7\%) 
$\mathcal{F}(1) = 0.919^{+ 0.030}_{-0.035}$~\cite{ref:auno}.  
This value is compatible with estimates based on non-lattice methods~\cite{ref:f1_1,ref:f1_2}.

\subsection{Form-factor parameterization}

Since HQET does not predict the functional form of the form factors, a
parameterization is needed for their extraction from the data. Perfect heavy quark symmetry implies that $R_1(w)=R_2(w)=1$, {\it i.e.}, the form factors $A_2$ and $V$ are identical for all values of $w$ 
and differ from $A_1$ only by a simple kinematic factor. 
Corrections to this approximation have been calculated in 
powers of $({\Lambda_{{\rm QCD}}}/{m_b})$ and the strong coupling constant $\alpha_s$.  Various parameterizations in powers of $(w-1)$ have been proposed. 
Among the different predictions relating the coefficients of the higher order
terms to the linear term, we adopt the following expressions
derived by Caprini, Lellouch, and Neubert~\cite{ref:CLNpaper},
\begin{eqnarray}
\label{eq:Cap}
h_{A_{1}}(w) & = &
h_{A_{1}}(1)
\big[ 1-8\rho^{2} z+(53\rho^{2}-15)z^{2} \nonumber \\ 
& &  -(231\rho^{2}-91)z^{3}\big], \\ 
R_{1}(w) & = & R_{1}(1)-0.12(w-1)+0.05(w-1)^{2},  \\ 
R_{2}(w) & = & R_{2}(1)+0.11(w-1)-0.06(w-1)^{2},  
\end{eqnarray}
\noindent
where  $z=[\sqrt{w+1}-\sqrt{2}]/[\sqrt{w+1}+\sqrt{2}]$.

The three parameters $\rho^{2}$, $R_{1}(1)$, and
$R_{2}(1)$, cannot be calculated; they must be extracted from data.

\section{DATA SAMPLE, RECONSTRUCTION, AND SIMULATION}
\label{sec:babar}

The data used in this analysis were recorded by the 
\babar\ detector, operating at the PEP-II asymmetric-energy
\epem\ collider.  The data sample corresponds to a luminosity of  
79~\invfb\ recorded on the \FourS\ resonance 
(on-resonance sample), and 9.6~\invfb\ recorded at 
a center-of-mass energy 40~MeV lower (off-resonance sample). 

The \babar\ detector and event reconstruction have been described in
detail elsewhere~\cite{ref:babar,ref:reconst}.  The momenta of charged
particles are measured by a tracking system consisting of a five-layer
silicon vertex tracker (SVT) and a 40-layer drift chamber (DCH).
Charged particles of different masses are distinguished by their
energy loss in the tracking devices and by a ring-imaging Cerenkov
detector (DIRC). Electromagnetic showers from electrons and photons
are measured in a CsI(Tl) calorimeter (EMC).  These detector
components are embedded in a $1.5$-$\mathrm{T}$ magnetic
field of the solenoid.  Electron candidates are selected on the basis of the ratio of
the energy detected in the calorimeter to the track momentum, the
calorimeter shower shape, the energy loss in the drift chamber, and
the angle of the photons reconstructed in the DIRC.  Muons are
identified in a set of resistive plate chambers inserted in the steel
flux-return of the magnet (IFR).  Information from the IFR is combined
with the track momentum measurement and energy deposition in the EMC
to improve the separation of muons from charged hadrons.

The electron and muon identification efficiencies and the
probabilities to misidentify a pion, kaon, or proton as an electron or
muon have been measured as a function of the laboratory momentum and
the angles with clean samples of tracks selected from
data~\cite{ref:thorsten}.
Within the acceptance of the calorimeter, defined by the polar angle
in the laboratory frame, $-0.72 < \cos \theta_{\mathrm{lab}} < 0.92$, and above
1.0 \gev, the average electron efficiency is $91\%$, largely
independent of the electron momentum.  The average hadron
misidentification rate is less than 0.2\%. The muon detection extends
to polar angles of $-0.91 < \cos \theta_{\mathrm{lab}} < 0.95$.  For a hadron
misidentification rate of typically 2.0\%, the average efficiency is
close to 65\%.

The criteria for distinguishing charged kaons from charged pions 
are chosen to maximize the efficiency while controlling the  
background, and thus differ for the decay modes under study. Consequently, the efficiency varies from 87\% to 97\%.  The uncertainties are typically 1\%.

We determine the tracking efficiency for high-momentum tracks by
comparing the independent information from SVT and DCH. We compute the
efficiency for low-momentum tracks reconstructed in the SVT alone from
the angular distribution of the ``slow" pion, \psoft, in the \Dstarm\ rest frame. We
use a large sample of $\Dstarm\ra\Dzb\psoft$, $\Dzb\ra \Kp\pim$ decays
selected from hadronic $B$ decays. For fixed values of the \Dstarm\
momentum, we compare the observed angular distribution to the one
expected for the decay of a vector meson to two pseudoscalar
mesons. 
This study is performed in several bins of the polar angle.
We define the relative efficiency as the ratio of the
observed to the expected distribution and parameterize its dependence
on the laboratory momentum of the \psoft. Below 100 \mev , this efficiency drops very steeply, and reaches zero at about 60 \mev. 

Neutral pions are reconstructed from pairs of photon candidates of
more than 30~\mev detected in the EMC and assumed to originate from
the interaction point. For photon pairs with an invariant mass within
15.75~\mev of the nominal \piz\ mass, we perform a kinematic fit,
constraining the mass. We require that the probability of the fit
exceeds 1\%.  The efficiencies, including the EMC acceptance, vary between 55\%
and 65\% for \piz\ energies ranging from 0.3 to 2.5 \gev, for a mass
resolution between 5.5 and 7.5~\mev.

We use Monte Carlo (MC) simulation of the production and decay of $B$
mesons at the \FourS\ resonance and of the detector
response~\cite{ref:geant4} to estimate signal and background
efficiencies, and to extract the observed signal and background
distributions to fit the data. We assume that the \FourS\ decays
exclusively to \BB\ pairs.  The simulated sample of generic \BB\
events corresponds to roughly three times the \BB\ data sample.

Information from studies of selected control data samples on
efficiencies and resolutions is used to improve the accuracy of the
simulation.  Comparisons of data with the MC simulations have revealed
small differences in the tracking and particle detection efficiencies,
which have been corrected for.  The MC simulations include radiative
effects such as bremsstrahlung in the detector material. QED 
final state radiation are modeled by PHOTOS 
\cite{ref:photos}, and decays with radiative photons are 
are included in the signal sample.

In the MC simulations the branching fractions for hadronic
$B$ and $D$ decays are based on average values reported in the
Review of Particle Physics~\cite{ref:pdg07}.  
$\Bz \ra \dsm \ellp \nu_{\ell} $ signal events are generated with the
HQET-based form factors, using the specific parameterization by
Caprini, Lellouch and Neubert~\cite{ref:CLNpaper}. Values of the 
form-factor parameters are taken from measurements by the CLEO
Collaboration~\cite{ref:CLEOff}.
$\B \to D^{**} \ell \nu$ decays, involving orbitally excited
charm mesons, are generated according to the ISGW2
model~\cite{ref:IGSW}, and decays to non-resonant charm states are
generated following the prescription of Goity and Roberts~\cite{ref:Goity}.

\section{EVENT SELECTION AND BACKGROUND SUBTRACTION}
\label{sec:recoandsel}

\subsection{Event selection}

The reconstruction of the events and the selection of candidate
\BztoDslnu\ decays are largely common to the earlier \babar\
analysis~\cite{ref:bad776}.

We select events that contain a \Dstarm~candidate and an oppositely
charged electron or muon with momentum in the range $1.2<p_{\ell}<2.4~\gev$.
Unless explicitly stated otherwise, momenta are measured in the
\FourS\ rest frame, which is boosted relative to the laboratory frame,
$\beta\gamma = 0.56$.
We reconstruct \Dstarm\ in the decay channel $\Dstarm \ra \Dzb \psoft$, 
with the \Dzb\ decaying to $\Kp \pim,~\Kp \pim \pip\pim$, or $\Kp \pim\piz$. 
The tracks of the charged hadrons from the \Dzb\ candidate are fitted 
to a common vertex and the candidate is rejected if the fit probability 
is less than $0.1\%$. We
require the invariant mass of the hadrons to be compatible with the
\Dzb\ mass within $\pm 2.5$ times the experimental resolution, corresponding to  $\pm 34\mev$ 
for $\Dzb \ra \Kp \pim\piz$ decays and $\pm 17\mev$ for the other decays.
For the decay $\Dzb \ra \Kp \pim\piz$, we accept only candidates from
portions of the Dalitz plot where the square of the decay
amplitude~\cite{ref:Dalitz} exceeds 10$\%$ of the
maximum.  
We select \Dstarm\ candidates with a momentum in the range
$0.5 < p_{D^*} < 2.5~\gev$.
For the \psoft\ from the
\Dstarm\ decay, the momentum in the laboratory frame must be less
than 450~\mev, and the momentum transverse to the beam must be greater than 50~\mev .
Finally, the lepton, the $\psoft$, and the \Dzb\ are fitted to a common
vertex using a constraint from the beam-beam interaction point. The probability for this fit is required to exceed 1\%.

In semileptonic decays, the presence of an undetected neutrino
complicates the separation of the signal from background. For 
a signal \Bz\ decay, the \Dstarm\ and the charged lepton originate 
from the \Bz\ and the only missing particle is a massless neutrino. 
The absolute value of the $B$ momentum, $p_B$, is known from the total 
energy in the event and its direction is constrained to lie
on a cone centered on the $\dsm \ellp$ momentum vector.  The opening angle
of this cone, $\TBY$, is computed for each event,
\begin{equation}
 \cos\TBY = \frac{2E_{B} E_{D^*\ell} - m^2_{B} - m^2_{D^*\ell} } { 2 p_{B} \; p_{\raisebox{-0.3ex}{\scriptsize $D^*\ell$}} }.
\end{equation}
Here $m, E$,  and $p$ refer to the mass, the energy, and the absolute value
of the momentum. The condition $|\cos\TBY|
\le 1.0$ should be fulfilled in a perfectly reconstructed decay.
The value of $w$ depends on the azimuthal angle of the \Bz\ direction,
which cannot be determined. We therefore approximate $w$ by
the average of  the four values of $w$ corresponding to the
azimuthal angles 0, $\pi/2$, $\pi$, and $3\pi/2$, as was done
in~\cite{ref:bad1224}. This approximation results in an average resolution
for $w$ of 0.04.

\subsection{Background subtraction}

The background subtraction is performed separately for each of the four kinematic variables ($w$, \ctl, \ctv, and $\chi$), to be used for the extraction of the form-factor parameters and \Vcb.  For each variable, we divide the events 
into ten subsets, each corresponding to one of ten bins of the distribution, 
and the fits are performed separately for the ten subsets.

The selected events are divided into six signal samples by separating
decays into electrons and muons and the three \Dzb\ decay modes.  In
addition to signal events, each subsample contains background events
from six different sources:
\begin{itemize}
\item {\it combinatorial background} (events from $\BB$ and continuum
      $q\bar{q}$ production in which at least one of the hadrons
      assigned to the \dsm\ does not originate from the
  \dsm\ decay);
\item {\it continuum background} ($\dsm \ellp$ combinations from $\epem \ra
  c\bar{c}$);
\item {\it fake lepton background}(a true \dsm combined with a hadron misidentified as a lepton);
\item {\it uncorrelated background} ($\ellp$ and \dsm\ originating from the decay
  of two different $B$ mesons);
\item \B\ {\it background involving higher mass charm states},  
  either $\B^+ \rightarrow \Dstar X \ellp \nul $ decays
  (via $\B^+ \rightarrow \overline{D}^{**0} \ellp \nul$ or
  non-resonant $\B^+ \rightarrow \Dstarm \pi^+ \ellp \nul$ charm states), 
  or $\Bz\ra\Dstarm X \ellp \nul $ decays, (via $\Bz\ra D^{**-} \ellp \nul$ 
  or non-resonant  $\Bz\ra\Dstarm \piz \ellp \nul $ charm states);
\item {\it correlated background events} due to the processes $\Bz \ra
  \dsm \tau^+ \nu_{\tau} $ with $\tau^+ \ra \ellp X$, and $\Bz \ra \dsm X_c$ with
  $X_c \ra \ellp Y$.

\end{itemize}

Except for the combinatorial background, all background sources
contain a true $\dsm \ra \Dzb \psoft$ decay and thus are expected to
exhibit a peak in the $\dm = m_{\dsm} - m_{\Dzb}$ distribution, where
$m_{\dsm}$ and $m_{\Dzb}$ are the reconstructed masses of the \dsm\
and \Dzb\ candidates.  We determine the background distributions from
data, except for the correlated background, which amounts to less than
1.8\% of the total sample of selected candidates.

\begin{figure}[htp]
\begin{center}
\scalebox{0.9}{\includegraphics{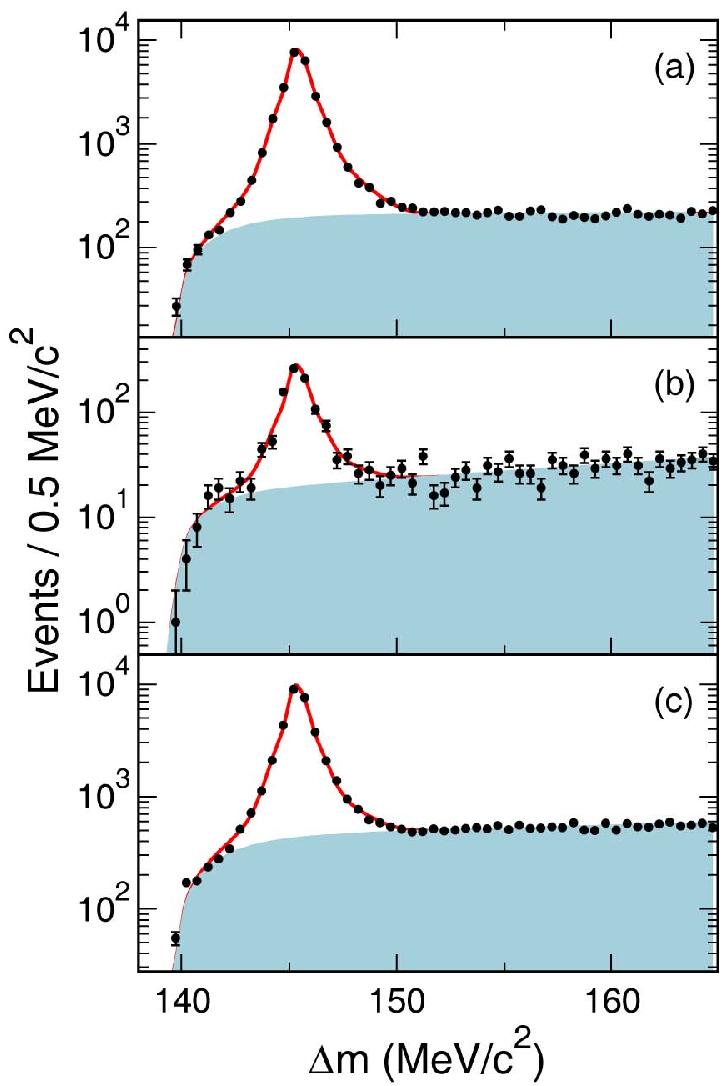}}
\end{center}
\caption{(color online) \dm\ distributions for the decays $\Dstarm \to
\Dzb \psoft$ with $\Dzb \to \Kp \pim$ for events in which the \psoft\
track is reconstructed in the SVT only, a) for on-resonance, b) for
off-resonance data in which an electron or muon was identified, and c)
for on-resonance data in which no charged lepton was found in the
event.  The data (points with statistical errors), integrated over the
full $w$ range, are compared to the result of the simultaneous fit
(solid line) to all three distributions. The shaded area shows the
fitted combinatorial background, the remainder (white area) represents
the sum of the signal and peaking backgrounds. }
\label{fig:dm}
\end{figure}

\begin{figure*}[htp]
\begin{center}
\scalebox{0.9} {\includegraphics{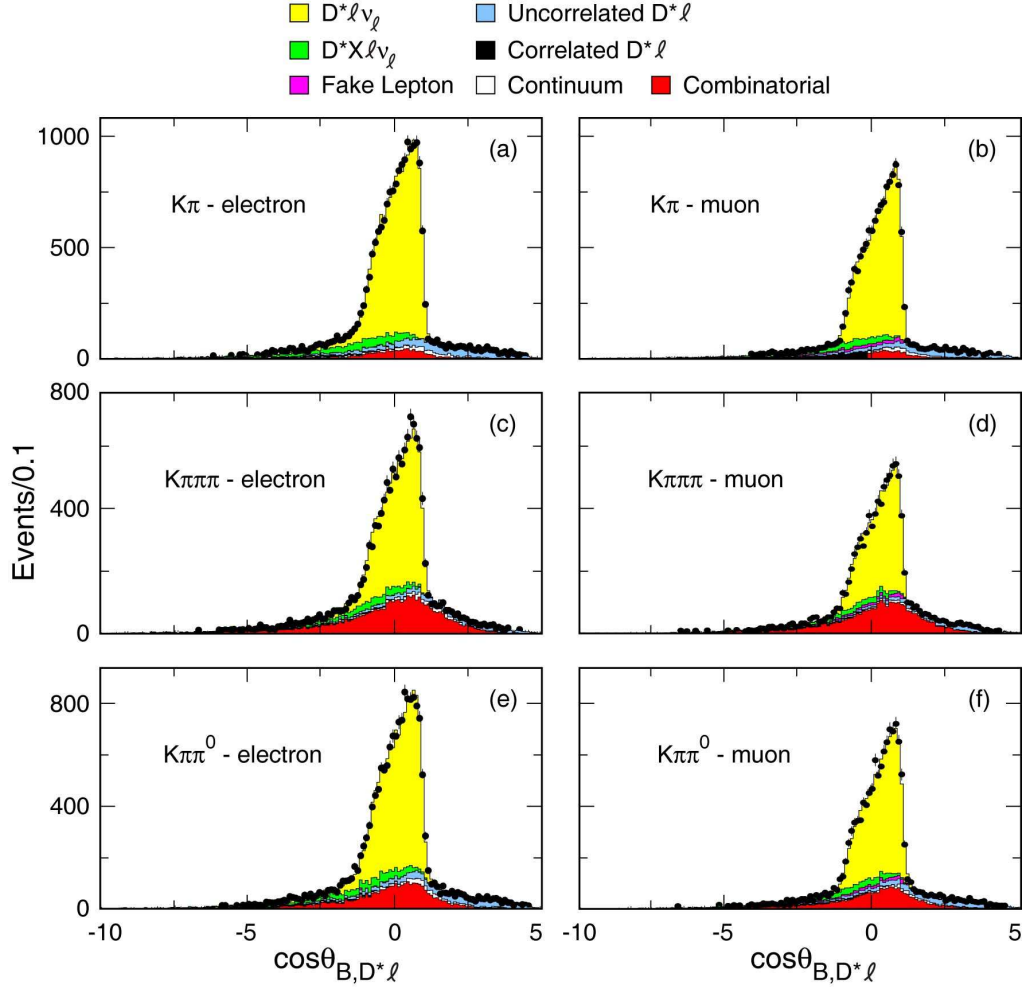}}
\end{center}
\vspace*{-0.4cm}
\caption{
The $\cos\TBY$ distributions (points with statistical errors) for selected events in the six subsamples (integrated over the full range in $w$) a,b) $D \to K \pi$, 
c,d) $D \to K \pi \pi \pi$, and e,f) $ D \to K \pi \pi^0$, compared to the result of the fits to signal and background contributions.
}
\label{fig:ctBY}
\end{figure*}

We determine the signal and background composition 
in two steps. 
First, we estimate the combinatorial, the continuum, and the fake lepton 
background from fits to the \dm\ distributions (Figure~\ref{fig:dm}).
Second, we fix the
background levels for these three sources and determine the uncorrelated
background and the $B \rightarrow \Dstar X \ell \nul $ background from
fits to the $\cos \TBY$ distributions (Figure~\ref{fig:ctBY}).
The shape and normalization of the 
small correlated background is fixed to the MC predictions and based on 
measured branching fractions~\cite{ref:pdg07}.

\subsubsection{Fits to $\dm$ distributions} 
To estimate the shape and normalization of the combinatorial, the
continuum, and fake lepton backgrounds from the measured \dm\
distributions, we use in addition to the on-resonance data,
off-resonance data, as well as a set of on-resonance events in which
no lepton is identified and a charged hadron is selected to take its
place. We refer to this data sample as the fake-lepton sample.
For each of these data sets, we perform an unbinned maximum likelihood
fit to the \dm\ distributions. The data are fitted to a sum of a peak,
due to correctly reconstructed $\dsm \ra \Dzb \psoft$ decays, and a
combinatorial background.

The peak is described as a sum of three Gaussian resolution functions
with three mean values, three different widths, and two parameters
that specify the contributions of continuum and fake lepton
backgrounds relative to the total number of events above the
combinatorial background.  The combinatorial background is described, as in~\cite{ref:bad776},
by an empirical function,
\begin{equation}
  \label{eq:combshape}
  F_{comb}(\dm)= \frac{1}{N}\, \left[ 1- e^{\left(-\frac{\dm
      - \dm_0}{c_1}\right)}\right]\,
\left(\frac{\dm}{\dm_0}\right)^{c_2} ,
\end{equation}
where $N$ is the normalization, $\dm_0$ refers to the kinematic threshold equal to the pion mass, and $c_1$ and $c_2$ are free parameters.

Since the \dm\ resolutions and background yields depend on the $\Dzb$
decay mode, and on whether the low-momentum pion track is
reconstructed in the SVT alone or in both SVT and DCH, the parameters
describing the peak contributions are determined separately for the
three subsamples corresponding to the $\Dzb$ decay modes, each divided
into two classes of events, depending on the detection of the slow
pion.

The off-resonance distributions are scaled to the on-resonance
luminosity, and the simulated lepton signal and fake lepton samples are
adjusted to reproduce the detection efficiencies and misidentification
probabilities determined from independently selected data control
samples.  For the on- and off-resonance data and for the fake-lepton
sample, the parameters describing the shape and normalization of
the combinatorial background differ, and they are therefore determined
separately, while the mean and the widths of the Gaussian functions
are common.

The fits extend in \dm\ from 138 to 165 \mev. Given the very large
number of parameters and the many data subsamples and various subsets,
the fits to the \dm\ distributions are performed in several
steps. Initially, the parameters describing the combinatorial
background contributions are fixed to values determined from binned
$\chi^2$ fits to simulated distributions.  With these starting values
for the combinatorial background shapes, an unbinned maximum
likelihood fit is performed to all subsets of the data, both on- and
off-resonance, to determine the eight peak-shape variables and three
additional relative peak yields for each data subset.  To improve the
agreement with the data, the parameters describing the combinatorial
background distributions are then refitted, together with the three
peak yield parameters, with the remaining parameters describing the
components of the peaking signal and background fixed to the results
of the previous fit.

As an illustration for these background fits, Figure~\ref{fig:dm} shows
the \dm\ distributions for the decays $\Dstarm \to \Dzb \psoft$ with
$\Dzb \to \Kp \pim$ for events in which the \psoft\ track is
reconstructed in the SVT only. The data are compared to the results of
the combined fit to three distributions, for selected samples of
on-resonance data, off-resonance data, and on-resonance data without an
identified lepton.  

The fitted fractions of combinatorial, fake-lepton, and
continuum background events are determined for the peak regions in \dm, which are defined as 144 to 147 \mev for
events with the slow charged pion detected in the SVT and DCH, and 143 to 148 \mev  for decays with the \psoft\ detected in the SVT alone.

\subsubsection{Fits to $\cos\TBY$ distributions} 

In a second step, a binned $\chi^2$ fit is performed to the $\cos\TBY$
distributions in the range $-10 < \cos\TBY < 5$, to determine the
signal contribution and the normalization of the uncorrelated and $B
\rightarrow \Dstar X \ell \nul $ background events.  Neglecting
resolution and radiative effects, signal events meet the constraint
$|\cos\TBY|\leq1$, while the distribution of $B \rightarrow \Dstar X
\ell \nul $ events extends below $-1$, and the uncorrelated background
events are spread over the entire range considered.
Because of final state radiation, the events in the electron samples also contribute to lower values of $\cos\TBY$.

This fit is performed separately for the six signal samples,
corresponding to the \Dzb\ mesons reconstructed in three different
decay modes, divided into events with electrons or muons. Each signal
sample is further divided into ten subsets corresponding to ten bins
in one of the four kinematic variables. We perform the fits separately
for each bin, with the individual shapes for the signal and for each
of the six background sources taken from MC simulation.
The fraction of the combinatorial, fake-lepton, and continuum events
are taken from the \dm\ fits and fixed.

To reduce the sensitivity to statistical fluctuations, we require that
the ratio of $B \rightarrow \Dstar X \ell \nul $ background and of
uncorrelated background to the signal be the same for all three \Dzb\
decay modes, with either electrons or muons.  The $\cos\TBY$
distributions for the six signal samples and the results of the fits
are shown in Figure~\ref{fig:ctBY}.  In total, there are 68,840 decay
candidates in the range $|\cos\TBY|<1.2$. The number of selected
events and the fractions of background events are given in
Table~\ref{tab:back}.
As expected, the fake rate is about a factor ten higher for decays to muons than to electrons.  Except for the
combinatorial backgrounds, which vary significantly for the three \Dzb\
decays modes, most of the other background fractions are similar for the six data samples.

\begin{table*}[htb]
\caption{Number of selected candidates and background fractions,
separately for the subsamples identified by the \Dzb\ decay mode and the charged lepton. 
}
\begin{center}
\begin{tabular}{lcccccc}
\hline \hline
Subsamples      & $K\pi \, e$ & $K\pi \, \mu$ & $K\pi\pi\pi \, e$ & $K\pi\pi\pi \, \mu$ & $K\pi\pi^0 \, e$ & $K\pi\pi^0 \, \mu$  \\
Number of Selected Candidates     & $15144$ & $12083$ & 
$10259$ & $7831$ & $13224$ & $10299$ \\
Signal Fraction ~[\%] & $85.27 \pm 0.31$ & $83.27 \pm 0.37$ & $69.25
\pm 0.54$ & $66.30 \pm 0.66$ & $75.72 \pm 0.43$ & $74.05 \pm 0.50$ \\
\hline
Background Source      &       &       & Background     & Fractions ~[\%]    &       &       \\
\hline
$B \to D^* X \ell \nu_{\ell}$ & $3.89 \pm 0.16$  & $ 3.93 \pm 0.18$  & $3.76 \pm 0.19$  & $3.74 \pm 0.22$  & $4.05 \pm 0.17$  & $4.08 \pm 0.20$  \\
Fake Leptons                  & $0.23 \pm 0.04$ &  $2.45 \pm 0.14$ & $0.12 \pm 0.03$    & $2.29 \pm 0.17$ & $0.22 \pm 0.04$ & $2.41 \pm 0.15$ \\
Uncorrelated                  & $3.59 \pm 0.15$ & $3.11 \pm 0.16$ & $2.71 \pm 0.16$     & $2.27 \pm 0.17$ & $3.38 \pm 0.16$ & $3.29 \pm 0.18$ \\
Correlated                    & $0.37 \pm 0.05$ & $0.56 \pm 0.07$ & $0.36 \pm 0.06$     & $0.48 \pm 0.08$ & $0.35 \pm 0.05$ & $0.47 \pm 0.07$ \\
Continuum                     & $1.80 \pm 0.11$ & $2.08 \pm 0.13$ & $1.82 \pm 0.13$     & $1.43 \pm 0.14$ & $1.83 \pm 0.12$ & $1.60 \pm 0.12$ \\
Combinatorial                 & $4.85 \pm 0.18$ & $4.60 \pm 0.20$ & $21.98 \pm 0.46$ & $23.49 \pm 0.55$ & $14.45 \pm 0.33$ & $14.10 \pm 0.37$ \\
\hline \hline
\end{tabular}
\end{center}
\label{tab:back}
\end{table*}

\section{EXTRACTION OF \boldmath\ensuremath{\Vcb} AND FORM-FACTOR PARAMETERS}
\label{sec:analysis}

We determine ${\cal F}(1)$\Vcb  and the three form-factor parameters
by extending the one-dimensional least-squares fit to the $w$ distribution
used previously~\cite{ref:bad776} to a combined fit of three
one-dimensional binned distributions, with bin-by-bin background
subtraction. We have chosen this approach to avoid the
statistical limitations of a fit to a binned four-dimensional decay
distribution.

In principle, any kinematic observable that is sensitive to the form-factor parameters can be used.  
We have examined the sensitivity of the four kinematic variables, $w$,
$\ctl$, $\ctv$ and $\chi$ (see Sec.~\ref{sec:formalism}), to the 
form-parameters and found that the $\chi$ distribution is
practically insensitive, and thus we select the remaining three
variables.
As done for the background estimate, we again divide the distributions into ten bins with equal bin size
for $w$ and $\ctv$, and varying bin size for $\ctl$ in order to have a more similar
population in all the bins for this variable.

We perform a least-squares fit to these three projected
one-dimensional distributions to extract the form factors and 
${\cal F}(1)$\Vcb,
using the independently-determined background estimates.  We account
for the correlations by noting that the statistical covariance between the content
of two bins in two different one-dimensional distributions is
determined by the common number of events in these bins, while it is
zero for bins in the same distribution.  Since each of the three
distributions that are included in the fit contain the same events and
thus have the same normalization, we reduce the total number of bins
used in the fit by two, from 30 to 28.  The choice of the bins that
are left out is arbitrary, and it has been verified that the fit
result does not depend on this choice.

\subsection{The least-squares fit}

The concept for this fit is an extension of the one introduced in the previous \babar\ analysis~\cite{ref:bad776}. 
For a given bin with index $i$, $N_i^{\mathrm{data}}$ is the total
number of observed events, $N_i^{\mathrm{bkg}}$ is the estimated
number of background events, and $N_i^{\mathrm{MC}}$ refers to the
number of MC-simulated signal events. The fit function can be
written as
\begin{eqnarray}
  X^2 & = & \sum_{i = 1}^{n_{bin} = 28} \sum_{j = 1}^{n_{bin} = 28} \biggl (N_i^{data} - N_i^{bkg} - \sum_{k=1}^{N_i^{MC}} \mbox{W}_i^k
  \biggr )  \nonumber \\ 
  & \times & C_{ij}^{-1} \biggl ( N_j^{data} - N_j^{bkg} - \sum_{k=1}^{N_j^{MC}} \mbox{W}_j^k \biggr ), 
\label{eq:chi2}
\end{eqnarray}
where the indices $i$ and $j$ run over the 28 bins, and the index $k$
runs over all MC-simulated events, $N_i^{\mathrm{MC}}$, in bin
$i$. $\mbox{W}_i^k$ is a weight assigned to the $k$-th simulated
signal event in bin $i$ to evaluate the expected signal yield as a
function of the free parameters of the fit, and $C_{ij}$ is the
covariance matrix element for a pair of bins $i$ and $j$.

Each weight $\mbox{W}_i^k$ is the product of four weights,
$\mbox{W}_i^k = \mbox{W}_i^{ff,k} \mbox{W}^{\mathcal{L}}
\mbox{W}_i^{\epsilon,k} \mbox{W}_i^{S,k} $. 
\begin{enumerate}
\item
The factor $\mbox{W}_i^{ff,k}$ accounts for the dependence of the
signal yield on the parameters to be fitted. The \Vcb dependence is
trivially given by the ratio $|V_{cb}|^2/|V_{cb}^{\mathrm{MC}}|^2$, where the
denominator is the actual value used in the simulation, derived from
the branching fraction assumed for the decay \BztoDslnu.  Similarly, the
dependence on the form-factor parameters, $\rho^2$, $R_1(1)$, and $R_2(1)$,
is given by the ratio of the differential decay rate
(Eq.~\ref{eq:totaldiffdecaywidth}), evaluated for the values assigned
during the fit and for the values adopted in the simulation.
\item
The factor $\mbox{W}^{\mathcal{L}}$ accounts
for the relative normalization of data and simulated samples. It is
the product of the following terms: 
\begin{itemize}
\item the ratio of the total number of $B\bar{B}$ events, $N_{B\bar{B}} = (85.9 \pm 0.9)
\times 10^6$ and the number of Monte Carlo events for the final
states $B^{0}\bar{B}^0$ and $B^+B^-$;
\item the ratios $1/[1+f_{+-}/f_{00}]$ for $B^0$,
   $[f_{+-}/f_{00}]/[1+f_{+-}/f_{00}]$ for $B^+$, where
  $f_{+-}/f_{00} = \mathcal{B}(\Upsilon(4S) \rightarrow
  B^+B^-)/\mathcal{B}(\Upsilon(4S) \rightarrow B^0\bar{B}^0)= 1.037
  \pm 0.029$
~\cite{ref:pdg07} is the ratio of number of charged to neutral $B$ mesons
  produced;
\item the ratio of the $c\bar{c}$ to $b\bar{b}$ pair-production cross section;
\item the ratio of the branching fraction
$\mathcal{B}(\Dstarm\rightarrow \Dzb \pim) = 0.677 \pm
0.005$~\cite{ref:pdg07} and of the $B^0$ lifetime $\tau_{B^0} = 1.530
\pm 0.009$ ps~\cite{ref:pdg07} to the values used in the Monte Carlo simulation.
\end{itemize}
The uncertainties on the measured quantities used in this weight function
are not accounted for in 
the fit; their impact is studied by repeating the fit with their
values changed by one standard deviation.
\item
The factor $\mbox{W}_i^{\epsilon,k}$ is the product of the
correction factors for efficiencies, applied on a particle by particle
basis, which accounts for the residual
differences in reconstruction and particle-identification efficiencies
between the data and the Monte Carlo simulation, as a
function of particle momentum and polar angle.
\item
The factor $\mbox{W}_i^{S,k}$ accounts for potential small differences
in efficiencies among the six data subsamples and allows the
adjustment of the $\Dzb$ branching fractions, properly accounting for
their correlated systematic uncertainties. $\mbox{W}_i^{S,k}$ is the
product of several scale factors that are free parameters in the fit,
each constrained to its expected value within the estimated
experimental uncertainty. Specifically, to account for the uncertainty in
the multiplicity-dependent tracking efficiency, we introduce a factor
$\mbox{W}_{trk}^{S} = 1 + N_{trk} (1 - \delta_{trk})$, where $N_{trk}$
is the number of charged tracks of the \Dsl candidates and
$\delta_{trk}$ is constrained to 1.0 within the estimated uncertainty
of $\sigma_{trk}=0.8$\% in the single-track efficiency. Similarly,
multiplicative correction factors $\delta_{\ell}$, $\delta_{K}$ and
$\delta_{\pi^0}$ are introduced to adjust the efficiencies for leptons
($\delta_{e}$ for $e^{\pm}$ or $\delta_{\mu}$ for $\mu^{\pm}$), kaons,
and $\pi^0$ mesons, each within their estimated uncertainties,
$\sigma_{\ell}$ ($\sigma_{e}$ for $e^{\pm}$ or $\sigma_{\mu}$ for
$\mu^{\pm}$), $\sigma_K,$ and $\sigma_{\pi}$.  Likewise,
$\delta_{{\cal B}(K\pi)}$, $\delta_{{\cal B} (K\pi\pi\pi)}$,
$\delta_{{\cal B}(K\pi\pi^0)}$ are introduced to adjust the individual
$\Dzb$ branching fractions within their current measurement
uncertainties~\cite{ref:pdg07}.  Correlations between the branching
fraction measurements are taken into account by the covariance matrix
$C_{{\cal B}}$.  As a result, $\mbox{W}_i^{S,k}$ can be expressed as
\begin{equation}
\nonumber 
\mbox{W}_i^{S,k} = \delta_{\ell}^{i,k}\delta_{K}^{i,k}(1 +
  N_{trk}^{i,k}(1-\delta_{trk}))\delta_{\pi^0}^{i,k}\delta_{{\cal B}}^{i,k} .
\label{eq:weight}
\end{equation}
The corrections to the kaon and $\piz$ efficiency, the decay
multiplicity, $N_{trk}$, and the $\Dzb$ branching fraction
depend on the particular event $k$ in bin $i$.
\end{enumerate}

The complete ansatz for the $X^2$ function used in the fit is
\begin{eqnarray}
  X^2 &=& \sum_{i = 1}^{n_{bin} = 28} \sum_{j = 1}^{n_{bin} = 28} \Bigg [
  \biggl (N_i^{data} -
  N_i^{bkg} - \sum_{k=1}^{N_i^{MC}} \mbox{W}_i^k \biggr )  \nonumber \\
  &\times& C_{ij}^{-1} 
  \biggl ( N_j^{data} - N_j^{bkg} -
  \sum_{k=1}^{N_j^{MC}} \mbox{W}_j^k \biggr )  \Bigg ] \nonumber \\
  & +& \frac{(1 - \delta_{\ell})^2}{\sigma_{\ell}^2} 
  + \frac{(1 - \delta_{K})^2}{\sigma_{K}^2} 
  + \frac{(1 - \delta_{trk})^2}{\sigma_{trk}^2} + \frac{(1 -
    \delta_{\pi^0})^2}{\sigma_{\pi^0}^2}  \nonumber \\
  & +& \sum_{m = 1}^3 \sum_{n = 1}^3 \delta_{{\cal B}( m)} 
     \times C_{{\cal B} (mn)}^{-1} \delta_{{\cal B} (n)} .
\label{eq:chi2full}
\end{eqnarray}
The indices $n,m$ refer to the three $\Dzb$ decay modes.
The addition of these extra terms allows us to fit all subsamples
simultaneously, while taking into account the correlated systematic
uncertainties and effectively propagating these uncertainties, via the
weights, to the uncertainties on the free parameters of the fit.

The fit procedure has been tested on a large variety of Monte Carlo
generated signal samples. The fits are performed for multiple samples,
comparable is size to the data. The resulting pulls are consistent with a 
Gaussian distribution, with no evidence for systematic biases. 
The width of the pull distribution has been found to be consistent with 1, 
giving confidence in the uncertainty extracted from the fit.

\subsection{The covariance matrix}
\label{sec:covariance}

A direct consequence of this method is that the covariance matrix for
the measurements is not diagonal, since the events in different bins
are not all statistically independent.  The total covariance matrix is
the sum of three separate matrices: one for the measured data yields,
and one each for the estimated signal and background yields.
The diagonal elements of the matrices are the uncertainties of the bin
contents. The covariance of bins belonging to the same distribution is
zero, and the covariance of bins from different distributions is the
variance of the number of events that is common for the two bins.

For the data, the covariance matrix is determined under the assumption
that the data
obey Poisson statistics, and therefore the variance
of the bin or of the intersection of two bins is simply the number of events. The estimated signal matrix is built in an
analogous way, where the variance of a number of weighted events $n$
is approximated by the sum of the squares of the weights, $\sum_{i = 1,n}
w_i^2$.

The calculation of the background covariance matrix is less
straightforward.  The diagonal elements are simply the estimated
variances of the measured background, according to the procedure
described in Sec.~\ref{sec:recoandsel}.  However, the background
extraction procedure does not directly determine the number of common
events in two bins because this procedure is based on a rather complex
sequence of fits to shapes and event yields.  The solution adopted
here is to use the number of common background events in two bins as
predicted by the simulation, corrected for tracking and particle
identification (PID) efficiencies and adjustments to account for the
background estimates.  This is done in such a way that for each bin of
one of the kinematic observables the background is equal to the
data-based background estimate.  The choice of the observable ($w,
\ctl, \ctv,$ or $\chi$) is arbitrary, and the variation of the results
with this choice is used to evaluate the systematic uncertainty
introduced by this procedure.

Since the total number of background events estimated for the four different
distributions is not exactly the same, we average the results obtained for the four distributions. The spread of the background normalization values is found to be almost twice as large as the estimated uncertainty from the 
error propagation. This can be explained by the fact that the
uncertainties of the assumed signal and background shapes in the
\dm\ fits are not accounted
for in this propagation.
We derive an estimate of this additional uncertainty using the number of background events determined from the fits to the background distributions. They are very similar for $w$ and $\ctl$, and for $\ctv$ and $\chi$. The estimate is given by the average of the minimum and maximum difference between these two groups.

\section{THE FIT RESULTS AND SYSTEMATIC UNCERTAINTIES}
\label{sec:results}

\subsection{Results of the fit}

The results of the simultaneous fit to the three one-dimensional distributions are presented in Table~\ref{tab:stdresu}. The stated uncertainties on 
$\rho^2, R_1(1), R_2(1)$ and $\mathcal{F}(1)|V_{cb}|$ are taken from the 
{\tt MINUIT} minimization program~\cite{ref:minuit}.
Among the form-factor parameters, the correlations are quite large, but their correlation with $\mathcal{F}(1)|V_{cb}| $ is less than 0.23.  
All the $\delta$ parameters are found to be consistent
with their nominal value of one within their uncertainties.

\begin{figure*}[htp]
\begin{center}
\scalebox{0.7} {\includegraphics{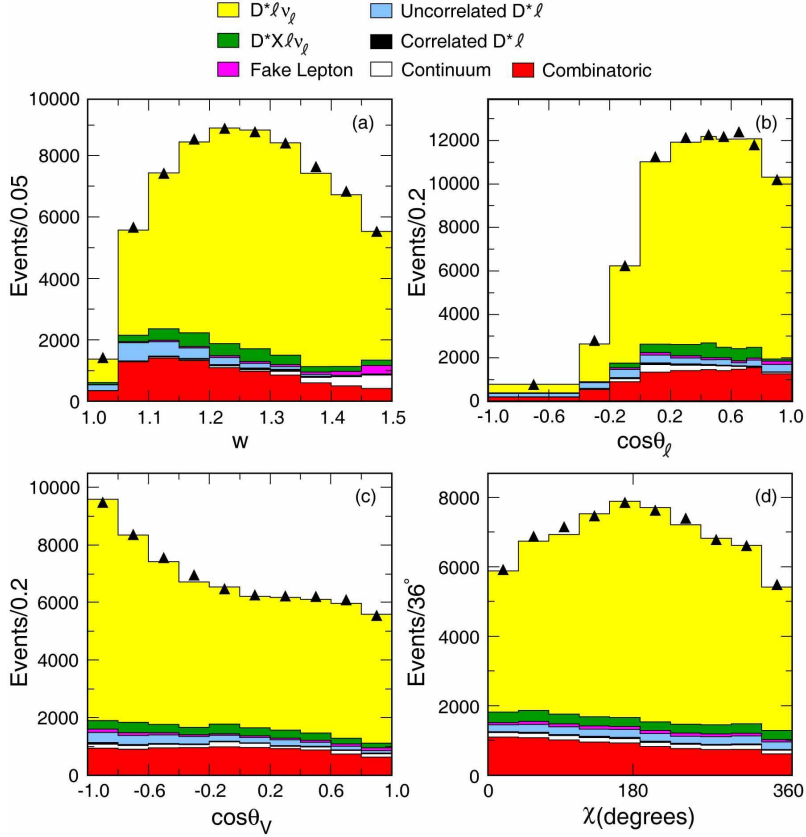}}
\end{center}
\vspace*{-0.4cm}
\caption{Comparison of the measured distributions (data points)   a) $w$, b) $\cos\theta_{\ell}$, c) $\cos\theta_{V}$, and d) $\chi$, with the result of the fit, shown as the sum of the fitted signal yield and the estimated background distributions. The statistical uncertainties of the data are too small to be visible.}
\label{fig:anchi_pnw}
\end{figure*}

Figure~\ref{fig:anchi_pnw} compares the one-dimensional projections for the kinematic observables with the results of the fit, detailing the signal and background contributions. To provide an additional check on the background estimation, we also show the $\chi$ projection. 
The goodness of the fit, ignoring that the measurements used are not
all independent, can be stated as  $\chi^2$/d.o.f. = 23.8/24,
corresponding to a probability of 47.3\%.  

\begin{table}[htb]
\caption{
The results of the fit: The parameters, their uncertainties, and the
off-diagonal elements of the correlation matrix. The stated uncertainties include
systematic ones that are not common to all events and are therefore included in the fit (see Sect.~\ref{sec:fiterrors}).}
\begin{center}
\begin{tabular}{lcccc}
\hline \hline 
           &   Fitted         &    & Correlations &    \\         
Parameters &   Values         & $\rho^2$ & $R_1(1)$ & $R_2(1)$    \\
\hline \noalign{\vskip2pt}
$\mathcal{F}(1)|V_{cb}| $   
           &$ (34.67 \pm 0.86) \times 10^{-3}$&$-0.001 $&$ -0.196 $&$ +0.141 $ \\
$\rho^2$   &$ 1.157 \pm 0.095              $&$       $&$ +0.867 $&$ -0.924 $ \\
$R_1(1)$   &$ 1.327 \pm 0.131              $&$       $&$        $&$ -0.928 $ \\
$R_2(1)$   &$ 0.859 \pm 0.077$&             &         &           \\
\hline \hline
\end{tabular}
\end{center}
\label{tab:stdresu}
\end{table}

The \BztoDslnu branching fraction, obtained by integrating the data over
the full phase space,
is found to be
$\mathcal{B}(\BztoDslnu) = ( 4.72 \pm 0.05 ) \%$.

\begin{table*}[htb]
\caption{
Results of fits performed separately for the six subsamples
corresponding to each combination of three \Dzb\ decay modes and the charged
lepton.  The uncertainties represent the total uncertainty of the fit, except for $\mathcal{F}(1)|V_{cb}|$, where it is split into the statistical and the systematic contribution included in the fit.
}
\begin{center}
\begin{tabular}{lccccc}
\hline \hline \noalign{\vskip2pt}
Subsample & $\rho^2$ & $R_1(1)$ & $R_2(1)$ & $\mathcal{F}(1)|V_{cb}| \times 10^3$ & $\chi^2$/d.o.f.\\
\hline
$K\pi \, e    $     & $0.971 \pm 0.163 $ & $1.166 \pm 0.182 $ & $0.977 \pm 0.107 $
                    & $34.76 \pm 0.61 \pm 0.61$ & 23.9/24 \\
$K\pi \, \mu  $     & $1.013 \pm 0.175 $ & $1.193 \pm 0.206 $ & $0.922 \pm 0.123 $
                    & $34.55 \pm 0.66 \pm 0.65$ & 37.9/24 \\
$K\pi\pi\pi \, e$   & $1.581 \pm 0.151 $ & $2.043 \pm 0.384 $ & $0.405 \pm 0.232 $ 
                    & $33.30 \pm 1.27 \pm 0.96$ & 15.6/24\\
$K\pi\pi\pi \, \mu$ & $1.146 \pm 0.258 $ & $1.156 \pm 0.351 $ & $0.946 \pm 0.197 $ 
                    & $34.14 \pm 1.10 \pm 0.98$ & 28.0/24\\
$K\pi\pi^0 \, e   $ & $1.042 \pm 0.165 $ & $1.217 \pm 0.206 $ & $0.926 \pm 0.118 $  
                    & $34.86 \pm 0.64 \pm 1.46$ & 26.9/24\\
$K\pi\pi^0 \, \mu $ & $1.170 \pm 0.155 $ & $1.439 \pm 0.228 $ & $0.838 \pm 0.131 $ 
                    & $34.38 \pm 0.74 \pm 1.46$ & 24.8/24\\

\hline \hline
\end{tabular}
\end{center}
\label{tab:elmu}
\end{table*}

As a cross check, we perform the fit separately for the six
subsamples. The quality of the fits is generally good, and the
results, shown in Table~\ref{tab:elmu}, agree within the uncertainties
obtained, with the possible exception of the $K \pi \pi \pi~ e$ sample.
Detailed investigations of the background estimates and fits for 
this sample did not reveal any anomalies.

\subsection{Systematic uncertainties}

A summary of statistical and systematic uncertainties on the measured parameters is presented in Table~\ref{table:totsys}, including the
breakdown of those for the measurement of the \BztoDslnu branching fraction.

\begin{table*}[ht]
\caption{
Breakdown of statistical and systematic uncertainties.}
\begin{center}
\begin{tabular}{lccccc}
\hline \hline \noalign{\vskip2pt}
        & $\rho^2$ & $R_1(1)$ & $R_2(1)$ & $\mathcal{F}(1)|V_{cb}| \times 10^3$ &
$\mathcal{B} (\BztoDslnu) \times 10^2$ \\
\hline
{\bf Statistical Error}
        &  {\bf 0.094}  &  {\bf 0.131}   & {\bf 0.077}    &{\bf 0.41}
        & {\bf 0.05} \\
\hline

PID, tracking, $\mathcal{B}(D^0)$  & 0.003 & 0.006 & 0.002 & 0.75 & 0.21 \\

Soft-$\pi$ efficiency         & 0.013 & 0.005 & 0.001 & 0.46 & 0.18 \\

$D^*l$ vertex fit     & 0.014 & 0.010 & 0.008 & 0.06 & 0.06 \\

$B$-momentum variation          & 0.013 & 0.040 & 0.017 & 0.29 & 0.14 \\

Radiative corrections         & 0.005 & 0.004 & 0.000 & 0.19 & 0.07 \\

$D^{**}$ composition
                   & 0.011 & 0.008 & 0.009 & 0.10 & 0.07 \\

Background estimates    & 0.006 & 0.004 & 0.002 & 0.04 & 0.04 \\

\hline

{\bf Partial Systematic Error}
               &{\bf 0.027}  &  {\bf 0.043}   & {\bf 0.021}    &{\bf
               0.95} & {\bf 0.33} \\
\hline

$B^0$ lifetime    & - & - & - & 0.10 & 0.03\\

$B\bar{B}$ normalization & - & - & - & 0.19 & 0.05 \\  

$\mathcal{B}(D^* \rightarrow D^0\pi)$ & - & - & - & 0.13 & 0.04 \\

$f_{+-}/f_{00}$ & 0.003 & 0.003 & 0.002 & 0.25 & 0.07 \\

\hline

{\bf Total Systematic Error}
               &{\bf 0.027}   &  {\bf 0.043}  & {\bf 0.021}    &{\bf 1.01} &{\bf 0.34} \\
\hline \hline
\end{tabular}
\end{center}
\label{table:totsys}
\end{table*}

\subsubsection{Uncertainties included in the fit}
\label{sec:fiterrors}
The uncertainty of the parameters resulting from the fit is not purely
statistical, since the systematic uncertainty sources that are not
common to all events are accounted for in the fit through the $\delta$
parameters in the weights $\mbox{W}_i^{S,k}$.
As described above, these weights account for residual uncertainties in the
lepton and hadron identification, the charged particle tracking and
$\pi^0$ efficiencies, and the individual $\Dzb$ branching fractions.
We can determine the statistical uncertainties of the fit by repeating the
fit with all $\delta$ parameters fixed at their fitted value.
The systematic uncertainties are then obtained by subtracting the statistical
covariance matrix
from the total covariance matrix.  While the uncertainties related to
the detector performance are relatively small for the form-factor
parameters, they are dominant for $\mathcal{F}(1)|V_{cb}|$
(2.5\%) and the branching fraction (5.2\%).

\subsubsection{Soft pion efficiency}

A major source of uncertainty on $\mathcal{F}(1)|V_{cb}|$ is the
reconstruction efficiency for the low-momentum pion from the \Dstarm\
decay, since it is highly correlated with the \Dstarm\ momentum and
thereby with $w$.  The functions parameterizing the efficiency for
data and MC simulation are consistent within the statistical
uncertainties. To assess the systematic uncertainty on \Vcb , we vary the
parameters of the efficiency function by their uncertainty, including
correlations. We add in quadrature the uncertainty in the absolute
scale as determined from higher-momentum tracks reconstructed in both
the SVT and the DCH.  The resulting systematic uncertainty on \Vcb\ is 1.3
\%.

\subsubsection{$D^*l$ vertex reconstruction efficiency}

The uncertainties from the $D^*l$ vertex reconstruction have been evaluated 
by observing the impact of changes in the standard vertex fit procedure.
First, we remove the constraint on the average position of the
beam-beam interaction point, and second, we remove the lepton track
from the vertex fit. We take the larger of the observed variations of
the parameters as the systematic uncertainty.

\subsubsection{$B$ momentum}

The $B$ momentum is impacted by small changes in the energies of the
two colliding beams. In the Monte Carlo simulation the beam energies
are assumed to be constant.  We have examined the impact of these variations on the measured distributions and have concluded that we can account for this difference by rescaling the $\cos\TBY$ values in the simulation by a factor of 0.97.
Half of the observed relative change of the fitted parameters coming
from the adjustment is assumed as the systematic uncertainty due to this effect.
This is the largest systematic uncertainty on $R_1(1)$ and $R_2(1)$.

\subsubsection{Radiative corrections}

Radiative corrections to the \BztoDslnu decays are simulated by {\tt
  PHOTOS 2.0}~\cite{ref:photos}, which describes the
final state photon radiation (FSR) up to $\mathcal{O}(\alpha^2)$.  In
the event reconstruction no attempt is made to recover photons emitted in the
decay. The simulated prediction of the
reconstructed kinematic variables is sensitive to the details of the radiative
corrections.  This is particularly important for electrons, for which FSR 
results in the long tail below $-1$ in the $\cos\TBY$ distribution.

At present, no detailed calculation of the full $\mathcal{O}(\alpha)$
radiative corrections to the \BztoDslnu decay is available.  Recently,
a new $\mathcal{O}(\alpha)$ calculation of radiative corrections in
$K^0 \to \pi^- e^+ \nu_e$ decays has become
available~\cite{ref:andre04}, which allows detailed comparisons of the
radiated photons with data and with {\tt PHOTOS} calculations. These
new calculations agree well with kaon data.
From the comparison with {\tt PHOTOS}, it is evident that the radiated
photon energy spectrum is quite well reproduced by {\tt PHOTOS}, but
the photon emission angle with respect to the electron differs
significantly.

To assess the systematic uncertainty due to the imperfect treatment of
the radiative corrections in $B$ decays, we have used the comparison
presented in Ref.~\cite{ref:andre04} and reweighted the simulated
decays to reproduce the photon angular distribution for photons above
10 \mev\ in the $B$ rest frame. The effect of the reweighting has been
used as an estimate of the systematic uncertainty. The possible impact
of the radiative effects on the hadronic interactions, {\em i.e.} the form
factors, is unknown and therefore not considered.

\subsubsection{$B \rightarrow \Dstar X \ell \nul $ background description}

The semileptonic branching fraction and form factors for the four
higher-mass charm states, $D^{**}$, and for non-resonant production of
$D\pi$ and $D^{*}\pi$ are not well known.
The shapes of these different components of the $B \rightarrow \Dstar
X \ell \nul $ background are taken from simulation. Their relative
yield is obtained from the fit to the $\cos\TBY$ distribution.
To account for the uncertainty of the composition of this background,
the fits have been repeated using only one of the contributions at a
time.
The study was done only for the $\Dzb \to \Kp \pim$ subsample; it is
assumed to be valid also for the other subsamples. The estimated uncertainty
on the fit parameters is taken as half of the biggest change observed.

\subsubsection{Background estimates}

As explained in Sec.~\ref{sec:covariance} the background covariance
matrix is built using the measured background shape in one observable,
and the simulation information for the others. The choice of the
observable is arbitrary. The systematic uncertainty is evaluated by
comparing results for the four kinematic observables. The maximum
observed variation with respect to the standard fit result is taken as
the estimate for this uncertainty.

\subsubsection{Global normalization factors}

There are several quantities that affect only the overall
normalization of the data, and thus not the form-factor
parameters. Their contribution to the systematic uncertainty on
$\mathcal{F}(1)|V_{cb}|$ and the branching fraction are listed in the
bottom half of Table~\ref{table:totsys}. They are: the $B^0$ lifetime
($\tau_{B^0} =
1.530 \pm 0.009$~\cite{ref:pdg07}), the $\Dstarm \to\Dzb \pim$
branching fraction ($\mathcal{B}(\Dstarm \to\Dzb \pim) = 67.7 \pm
0.5 \%$), and $N_{B\bar{B}}$, the number of $B\bar{B}$ events in the total data sample.
The systematic uncertainty on $N_{B\bar{B}}$ is 1.1\%.  

The effect of the uncertainties in the $\Dzb$ branching fractions has
already been discussed; it is subsample specific, but it affects all
parameters because it changes the fraction of signal events from
different $\Dzb$ decays.

The uncertainty on the 
ratio $f_{+-}/f_{00} = \mathcal{B}(\Upsilon(4S) \rightarrow B^+B^-)/\mathcal{B}(\Upsilon(4S) \rightarrow B^0\bar{B}^0)= 1.037 \pm 0.029$
~\cite{ref:pdg07} affects both the absolute number
of measured \BztoDslnu decays 
and the ratio of background events from $B^0$ and $B^{\pm}$ decays in
the simulation. This second aspect influences the $\cos\TBY$
distributions and therefore the background determination. For this
reason, this uncertainty
affects also the form-factor parameters. The systematic uncertainty is
equated with the observed change in the fit parameters for a one-standard deviation change in the value of $f_{+-}/f_{00}$.

\section{CONCLUSIONS}
\label{sec:summary}

\subsection{Summary of results of this analysis}

A sample of about 52,800 fully reconstructed \BztoDslnu decays
recorded by the \babar\ detector has been analyzed to extract both
$\mathcal{F}(1)|V_{cb}|$ and the form-factor parameters, $\rho^2$,
$R_1(1)$ and $R_2(1)$, in the Caprini-Lellouch-Neubert
parameterization~\cite{ref:CLNpaper}.
The \Dstarm candidates are reconstructed from the \Dstarm\ra\Dzb\pim\
decays and the \Dzb\ mesons are reconstructed in three different
decay modes, $\Kp\pim$, $\Kp\pim\pip\pim$, and
$\Kp\pim\piz$. Electrons or muons are paired with the \Dstarm\
to form signal candidates.  The large data sample has permitted a 
more precise determination of the background contributions, 
largely based on data, and thus has resulted in smaller experimental
uncertainties. 

The results of the simultaneous fit to three one-dimensional projections 
of the decay rate are 
\begin{eqnarray*}
  \mathcal{F}(1)|V_{cb}| & = & (34.7 \pm 0.4 \pm 1.0) \times 10^{-3} \\
  \rho^2 & = & 1.157 \pm 0.094 \pm 0.027 \\
     R_1(1) & = & 1.327 \pm 0.131 \pm 0.043 \\
     R_2(1) & = & 0.859 \pm 0.077 \pm 0.021 . 
\end{eqnarray*}
The stated uncertainties of the measurement here are the statistical and the total  
systematic one.  The simultaneous fit
to the three distribution reduces the uncertainty due to the form-factor parameters.

\begin{figure}[htp]
\begin{center}
\scalebox{0.5}{\includegraphics{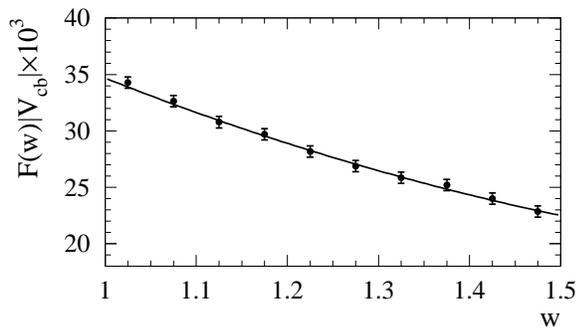}}
\end{center}
\vspace*{-1.0cm}
\caption{
The measured $w$ dependence of $\mathcal{F}(w)|V_{cb}|$ (data points) compared to the 
theoretical function with the fitted parameters (solid line).
The experimental uncertainties are too small to be visible.
}
\label{fig:fwvcb}
\end{figure}

Using an unquenched lattice calculation giving $\mathcal{F}(1) = 0.919^{+
0.030}_{- 0.035}$~\cite{ref:auno} results in the following value for
$|V_{cb}|$,
\begin{equation*}
  |V_{cb}| = ( 37.8 \pm 0.4 \pm 1.1 ^{+ 1.2}_{- 1.4} ) \times
   10^{-3}.
\end{equation*}
Here the third error is due to the theoretical uncertainty in $\mathcal{F}(1)$. 
Figure~\ref{fig:fwvcb} shows the measured decay rate, integrated over angles, 
$\mathcal{F}(w)|V_{cb}|$, as well as the fitted 
theoretical $w$ dependence (see Eq. \ref{eq:dgamma}).

The branching fraction for the decay \BztoDslnu is
\begin{equation*}
  \mathcal{B}(\BztoDslnu) = (4.72 \pm 0.05 \pm 0.34 ) \%.
\end{equation*}

\subsection{Combination of results with the previous \babar\ measurement of the form-factor parameters }
\label{sec:combi}
The \babar\ collaboration recently published a
measurement~\cite{ref:bad1224} of the same form-factor parameters for
\BztoDslnu\ decays based on an unbinned maximum-likelihood fit to the
four-dimensional decay distribution (Eq.
\ref{eq:totaldiffdecaywidth}).  This fit is sensitive to the
interference of the three helicity amplitudes and thus results in
significant smaller uncertainties on the form-factor parameters.  The fit
does not attempt an absolute normalization of the decays, and thus is
not sensitive to $\mathcal{F}(1)|V_{cb}|$.  It resulted in $\rho^2 =
1.145 \pm 0.066 \pm 0.035$, $R_1(1) = 1.396 \pm 0.070 \pm 0.027, $ and
$ R_2(1) = 0.885 \pm 0.046 \pm 0.013$.
 
We combine the two \babar\ measurements of the form-factor parameters, taking
into account the correlation between them, and obtain
\begin{eqnarray*}
\mathcal{F}(1)|V_{cb}| & = & ( 34.4 \pm 0.3 \pm 1.1) \times 10^{-3}  \\
\rho^2 & = & 1.191 \pm 0.048 \pm 0.028 \\
R_1(1) & = & 1.429 \pm 0.061 \pm 0.044 \\
R_2(1) & = & 0.827 \pm 0.038 \pm 0.022.
\end{eqnarray*}
Compared to the analysis presented in this paper, the combined result
has significantly smaller statistical uncertainties of the form-factor
parameters.
The event sample and the sample of Monte Carlo simulated events used
in Ref.~\cite{ref:bad1224} are a subset of the one used in the present
analysis, namely about 15,000 selected \Bzb\ candidates with $\Dzb \to
\Kp\pim$ decays combined with electrons. Except for the selection of
the $\Dzb$ decay, the event selection and the determination of the
backgrounds shapes and the signal extraction are almost identical for
the two analyses.  Therefore, all the detector-related systematic
uncertainties should be the same, as well as the uncertainties from the
background models and input parameters like the branching fractions.
Thus, we retain the systematic measurement uncertainties established in this
paper.  The combined statistical errors are still larger than the
total systematic uncertainties, but not by a large factor.
An upper limit for the correlation between the two measurements
has been estimated on the basis of the ratio of the uncertainties, and
is found to be less than $0.45$.
 
The correlation coefficients for the combined measurements are
\begin{eqnarray*}
\rho(\rho^2,R_1(1)) & = & +71\% \\
\rho(\rho^2,R_2(1)) & = & -83\% \\
\rho(\rho^2,\mathcal{F}(1)|V_{cb}|) & = & +27\% \\
\rho(R_1(1),R_2(1)) & = & -84\% \\
\rho(R_1(1),\mathcal{F}(1)|V_{cb}|) & = & -39\% \\ 
\rho(R_2(1),\mathcal{F}(1)|V_{cb}|) & = & +22\% .
\end{eqnarray*}

\begin{figure*}[htp]
\begin{center}
\scalebox{1.0} {\includegraphics{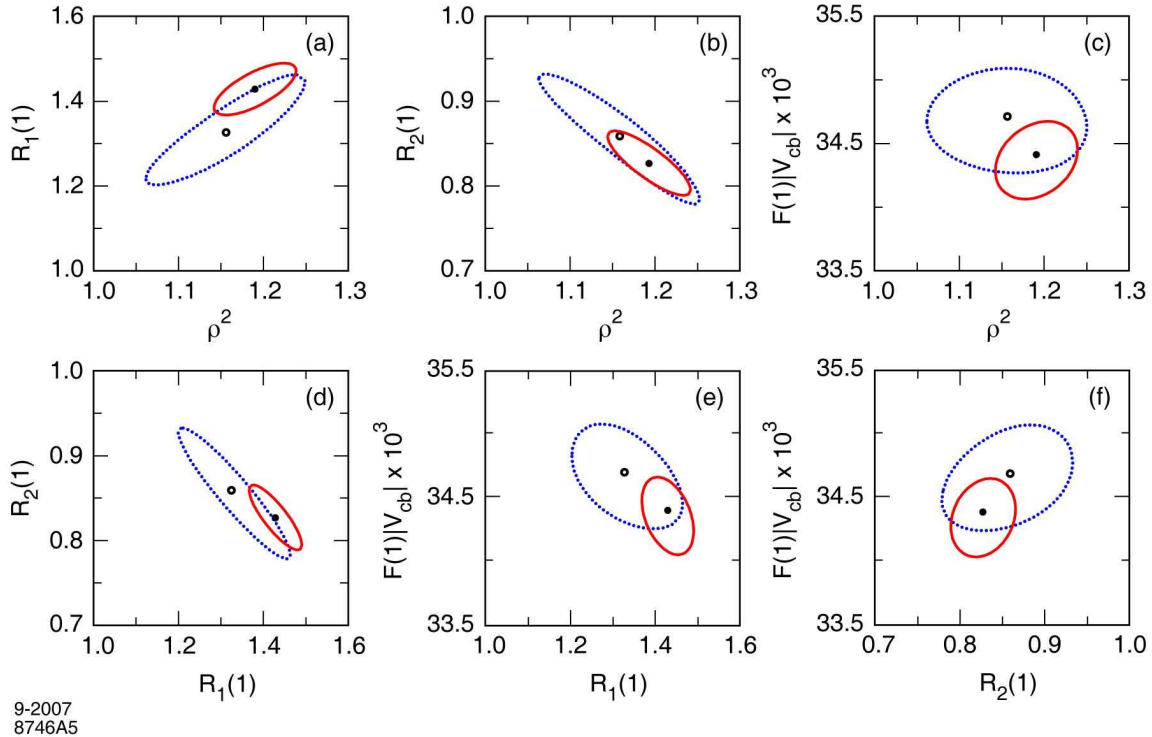}}
\end{center}
\vspace*{-0.4cm}
\caption{(color online)  Correlations between fitted variables and their
uncertainties, both for the present analysis with the statistical
uncertainties (dotted line, central value marked with an open circle) and for
the combined result with the total experimental uncertainties, {\em i.e.} the
statistical and systematic uncertainties combined (solid line, central value
marked with a solid dot),
a) $\rho^2$-$R_1(1)$, 
b) $\rho^2$-$R_2(1)$, 
c) $\rho^2$-$\mathcal{F}(1)|V_{cb}| \times 10^3$, 
d) $R_1(1)$-$R_2(1)$,    
e) $R_1(1)$-$\mathcal{F}(1)|V_{cb}| \times 10^3$, and
f) $R_2(1)$-$\mathcal{F}(1)|V_{cb}| \times 10^3$ projections. The contours correspond to $\Delta \chi^2 = 1$, {\em i.e.} 39\% CL.
}
\label{fig:correlations}
\end{figure*}

Figure~\ref{fig:correlations} shows the
correlations between the fitted variables and their 
uncertainties, both for the present analysis and for the combined
result with Ref.~\cite{ref:bad1224}.
The contours correspond to $\Delta \chi^2 = 1$, {\em i.e.} 39\% CL.
The correlations between the form-factor parameters are quite large,
but their correlation with $\mathcal{F}(1)|V_{cb}|$ is less than 0.4,
and the sign of the coefficients differ, resulting in a much reduced
overall dependence of $\mathcal{F}(1)|V_{cb}|$ on these form factors.

Using the same lattice calculation for $\mathcal{F}(1)$~\cite{ref:auno}, we
obtain an improved value for $|V_{cb}|$,
$$|V_{cb}| =(37.4 \pm 0.3  \pm  1.2 \, ^{+1.2}_{-1.4} ) \times 10^{-3},$$ 
\noindent
where the third error reflects the current uncertainty on  $\mathcal{F}(1)$.  

The corresponding branching fraction of the decay \BztoDslnu is
\begin{equation*}
  \mathcal{B}(\BztoDslnu) = (4.69 \pm 0.04 \pm 0.34 ) \%.
\end{equation*}   
The combined results of the two \babar\ analyses supersede 
all previous \babar\ measurements of the
form-factor parameters, of the exclusive branching fraction for the
\BztoDslnu decay, and of \Vcb extracted from this decay.

The value of the branching fraction presented here is smaller than the
average of previous measurements~\cite{ref:pdg07}. This measurement combined with
$\mathcal{B}(\Bz \ra \Dm \ell \nu_{\ell}) = (2.08 \pm
0.18)\%$~\cite{ref:pdg07} represents only $(65 \pm 7)\%$ of the
total branching fraction for the $\Bz \ra X_c \ell \nu_{\ell}$ decays.
The remaining fraction of 35\% is expected to involve higher-mass charm states.
The branching fractions for decays to these individual higher-mass states are not well
known, in particular those involving broad resonances or non-resonant
$D^{(*,**)} \pi$ states~\cite{ref:belle05,ref:pegna}.

The combination of the two \babar\ measurements results in a further reduction 
of the form-factor uncertainties compared to the previous \babar\
analysis~\cite{ref:bad1224}, for which the uncertainties on $R_1(1)$ and
$R_2(1)$ had already been reduced by a factor of four or more,
compared to the CLEO measurement~\cite{ref:CLEOff}.
The uncertainty on $\rho^2$ has also been reduced, by a factor of
five with respect to the \babar\
measurement in Ref.~\cite{ref:bad776}. The correlation between
$\mathcal{F}(1)\Vcb$ and $\rho^2$, which was sizable for all previous
measurements, has been reduced significantly, and this also leads to a
smaller uncertainty for $|V_{cb}|$.

The resulting value of $|V_{cb}|$ is fully compatible with the earlier \babar\
measurement~\cite{ref:bad776}, and most earlier measurements~\cite{ref:pdg07}, 
but it is significantly smaller than the CLEO measurement~\cite{ref:CLEOVcb}. 

\Vcb\ can also be extracted from inclusive $B \to X_c \ell \nu_{\ell}$ decays. 
Recent measurements of \Vcb\ rely on moments of the electron energy
and the hadron mass spectrum, combined with moments of the energy
photon spectrum in inclusive $B \to X_s \gamma$ decays.  Here $X_c$
and $X_s$ refer to charm  and strange hadronic states, resonant or
non-resonant. 
Global fits to such moments, as a function of the minimum lepton and
photon energy
have been performed in terms of two different QCD calculations, one in
the so-called 1S scheme~\cite{ref:bauer} and the other in the kinetic
scheme~\cite{ref:buchmuller}.  The results are 
in very good agreement. The weighted average is 
$|V_{cb}| =(41.7 \pm 0.7)\times 10^{-3}$~\cite{ref:pdg07}; the
experimental and theoretical uncertainties are comparable in size.   
The experimental techniques and the theoretical calculations for
exclusive and inclusive decays are completely different and
uncorrelated.  Given the sizable experimental error and the large
uncertainty of the form-factor normalization of the exclusive
measurement, the results are consistent.  This should give us
confidence in the experimental methods employed and the theoretical
calculations, but also an incentive to improve on both.

\vspace{0.5cm}
\section{ACKNOWLEDGMENTS}
\label{sec:Acknowledgments}

We are grateful for the 
extraordinary contributions of our \pep2\ colleagues in
achieving the excellent luminosity and machine conditions
that have made this work possible.
The success of this project also relies critically on the 
expertise and dedication of the computing organizations that 
support \babar.
The collaborating institutions wish to thank 
SLAC for its support and the kind hospitality extended to them. 
This work is supported by the
US Department of Energy
and National Science Foundation, the
Natural Sciences and Engineering Research Council (Canada),
the Commissariat \`a l'Energie Atomique and
Institut National de Physique Nucl\'eaire et de Physique des Particules
(France), the
Bundesministerium f\"ur Bildung und Forschung and
Deutsche Forschungsgemeinschaft
(Germany), the
Istituto Nazionale di Fisica Nucleare (Italy),
the Foundation for Fundamental Research on Matter (The Netherlands),
the Research Council of Norway, the
Ministry of Science and Technology of the Russian Federation, 
Ministerio de Educaci\'on y Ciencia (Spain), and the
Particle Physics and Astronomy Research Council (United Kingdom). 
Individuals have received support from 
the Marie-Curie IEF program (European Union) and
the A. P. Sloan Foundation.

   



\begin{thebibliography}{99}
\bibitem{footnote1} Charge conjugate decay modes are implicitly included, 
and $\ell$ refers to a charged lepton, either an electron or muon.
\end{thebibliography}
\end{document}